\documentclass[aps,prb,reprint,superscriptaddress]{revtex4-1}

\usepackage{graphicx}
\usepackage{dcolumn}
\usepackage{bm}
\usepackage{amsmath}
\usepackage{hyperref}

\begin{document}


\title{Magnetotransport in heterostructures of transition metal dichalcogenides and graphene}


\author{Tobias V\"olkl}
\affiliation{Institut f\"ur Experimentelle und Angewandte Physik, Universit\"at Regensburg, Germany}
\author{Tobias Rockinger}
\affiliation{Institut f\"ur Experimentelle und Angewandte Physik, Universit\"at Regensburg, Germany}
\author{Martin Drienovsky}
\affiliation{Institut f\"ur Experimentelle und Angewandte Physik, Universit\"at Regensburg, Germany}
\author{Kenji Watanabe}
\affiliation{NIMS, 1-1 Namiki, Tsukuba, Japan}
\author{Takashi Taniguchi}
\affiliation{NIMS, 1-1 Namiki, Tsukuba, Japan}
\author{Dieter Weiss}
\affiliation{Institut f\"ur Experimentelle und Angewandte Physik, Universit\"at Regensburg, Germany}
\author{Jonathan Eroms}
\email{jonathan.eroms@ur.de}
\affiliation{Institut f\"ur Experimentelle und Angewandte Physik, Universit\"at Regensburg, Germany}



\date{\today}

\begin{abstract}
We use a van-der-Waals pickup technique to fabricate different heterostructures containing WSe$_2$(WS$_2$) and graphene. The heterostructures were structured by plasma etching, contacted by one-dimensional edge contacts and a topgate was deposited. For graphene/WSe$_2$/SiO$_2$ samples we observe mobilities of $\sim$12\,000\,cm$^2$/Vs. Magnetic field dependent resistance measurements on these samples show a peak in the conductivity at low magnetic field. This dip is attributed to the weak antilocalization\,(WAL) effect, stemming from spin-orbit coupling. Samples where graphene is encapsulated between WSe$_2$(WS$_2$) and hBN show a much higher mobility of up to $\sim$120\,000\,cm$^2$/Vs. However, in these samples no WAL peak can be observed. We attribute this to a transition from the diffusive to the quasiballistic regime. At low magnetic field a resistance peak appears, which we ascribe to a size effect, due to boundary scattering. Shubnikov-de Haas oscillations in fully encapsulated samples show all integer filling factors, due to complete lifting of the spin and valley degeneracy.
\end{abstract}


\maketitle



%
\section{Introduction}
In recent years, the assembly of van-der-Waals heterostructures containing graphene has gained much attention\,\cite{Geim2013}. Encapsulating graphene between hBN and employing one-dimensional edge contacts\,\cite{Wang614} has proven to be a reliable method to fabricate high mobility devices. With this a number of  effects, such as ballistic transport\,\cite{Taychatanapat2013}, viscous electron flow\,\cite{Bandurin2016} and moir\'e patterns\,\cite{Dean2013} have been observed. However, employing other two-dimensional materials for encapsulation allows to further tailor the properties of graphene. One promising objective is to increase the spin-orbit-coupling\,(SOC) in graphene, as this may offer numerous possibilities, including the generation of a pure spin-current through the spin-Hall effect or the manipulation of spin-currents through an electric field. Bringing graphene into proximity of transition metal dichalcogenides\,(TMDC) has been predicted theoretically\,\cite{Kaloni2014,Gmitra2016} and observed experimentally\,\cite{Wang2015a,Wang2016,Yang2016,Avsar2014} to increase SOC in graphene. Further, transport measurements\,\cite{Kretinin2014} and recent Raman measurements indicate the suitability of these substrates for high mobility graphene\,\cite{Banszerus2017}. This is in contrast to previously explored methods for increasing SOC in graphene, such as hydrogenation\,\cite{CastroNeto2009,Balakrishnan2013a}, fluorination\,\cite{Avsar2015} or the attachment of heavy atoms\,\cite{Ma2012,Balakrishnan2014}, as these methods have the disadvantage of increasing the scattering and therefore decreasing the mobility of graphene.\\
Here, we report on a comparison of magnetotransport in graphene/TMDC heterostructures in a broad mobility range, realized by different material combinations in the van-der-Waals stacked layer sequence. We integrate one-dimensional contacts into the TMDC/graphene processing scheme, achieving a high yield of functional devices and include top gates using a TMDC layer as a gate dielectric. In diffusive samples, we observe weak antilocalization and study proximity-induced spin-orbit interaction at different out-of-plane electric fields, while in high mobility samples, a ballistic size effect and the quantum Hall effect are observed.\\
\section{Sample Fabrication}
\begin{figure}
	\begin{minipage}{0.44\columnwidth}
		\includegraphics[width=\columnwidth]{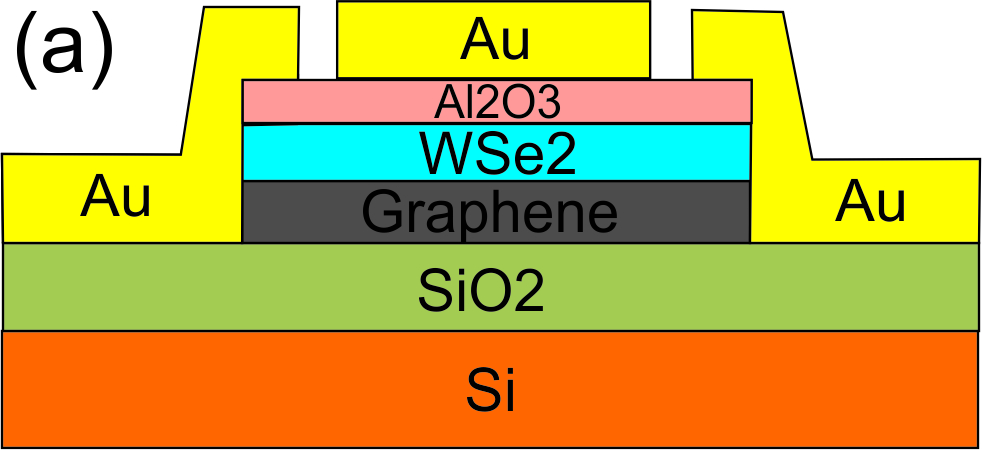}
		\includegraphics[width=\columnwidth]{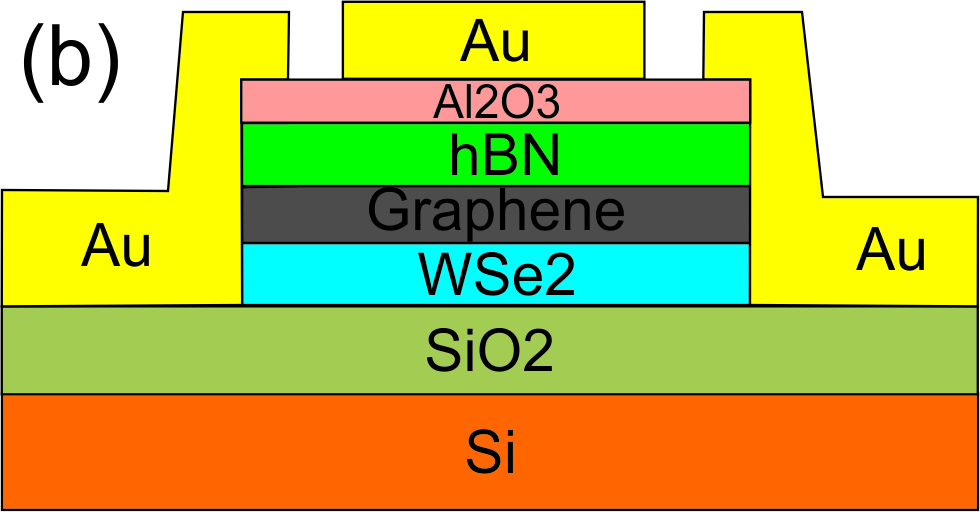}
	\end{minipage}
	\hfill
	\begin{minipage}{0.54\columnwidth}
		\includegraphics[width=\columnwidth]{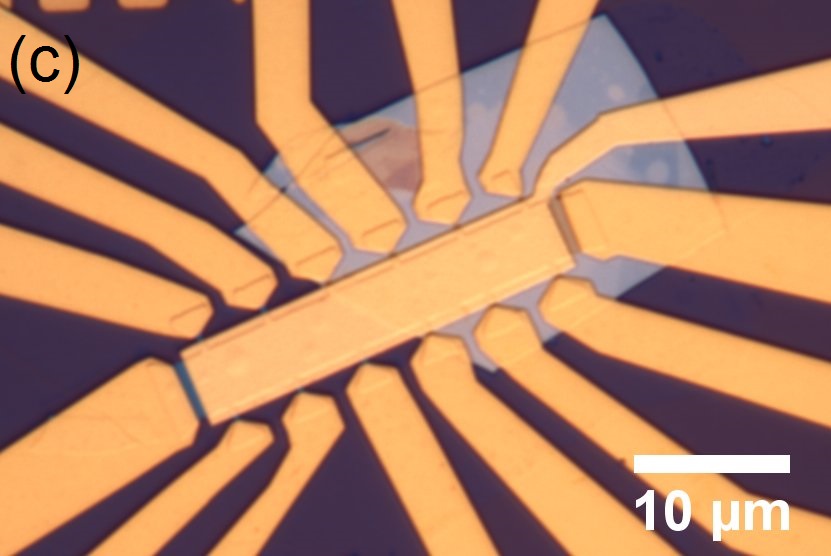}
	\end{minipage}
	\caption{(a,b) Schematic cross section of the devices. (a) Cross section of device type\,1, consisting of monolayer graphene and WSe$_2$ on top of SiO$_2$. (b) Cross section of device type\,3. Bilayer graphene is encapsulated between hBN and WSe$_2$. (c) Optical microscope picture of device 3. Part of the hBN/graphene stack lies on a WSe$_2$ flake, the other part lies directly on the SiO$_2$ substrate.\label{1}}
\end{figure}

Heterostructures were fabricated by using a dry pickup process\,\cite{Wang614}. Three different types of devices were fabricated. For device type\,1\,(see Fig. \ref{1}\,(a)) monolayer graphene was picked up by exfoliated multilayer WSe$_2$ and placed onto a standard p$^{++}$-doped Si/SiO$_2$ chip.
For device type\,2 monolayer graphene was encapsulated between hBN and WS$_2$, while for device type\,3\,(see Fig. \ref{1}\,(b)) bilayer graphene was encapsulated between hBN and WSe$_2$. After assembly all three devices were annealed for 1\,hour at 320\,$^{\circ}$C in vacuum and 1\,hour at 320\,$^{\circ}$C in forming gas. Annealing removes contaminations between the layers, as well as the remaining PPC on top of the WSe$_2$\,(WS$_2$) flake. Then electron-beam lithography\,(EBL) and reactive ion etching\,(RIE) with CHF$_3$/O$_2$ were employed to define a Hall-bar structure. The graphene was then contacted by 5\,nm Cr/ 80\,nm Au side contacts. These edge contacts showed high reliability as 70 of 74 contacts were functional. As a last step 10\,nm Al$_2$O$_3$ were deposited by atomic layer deposition\,(ALD), followed by a Au topgate. The Al$_2$O$_3$ layer is necessary to prevent any leakage between topgate and graphene at the sides of the stack.\\
\section{Experimental Results}
\subsection{Diffusive Regime}
\begin{figure}
	\includegraphics[width=\columnwidth]{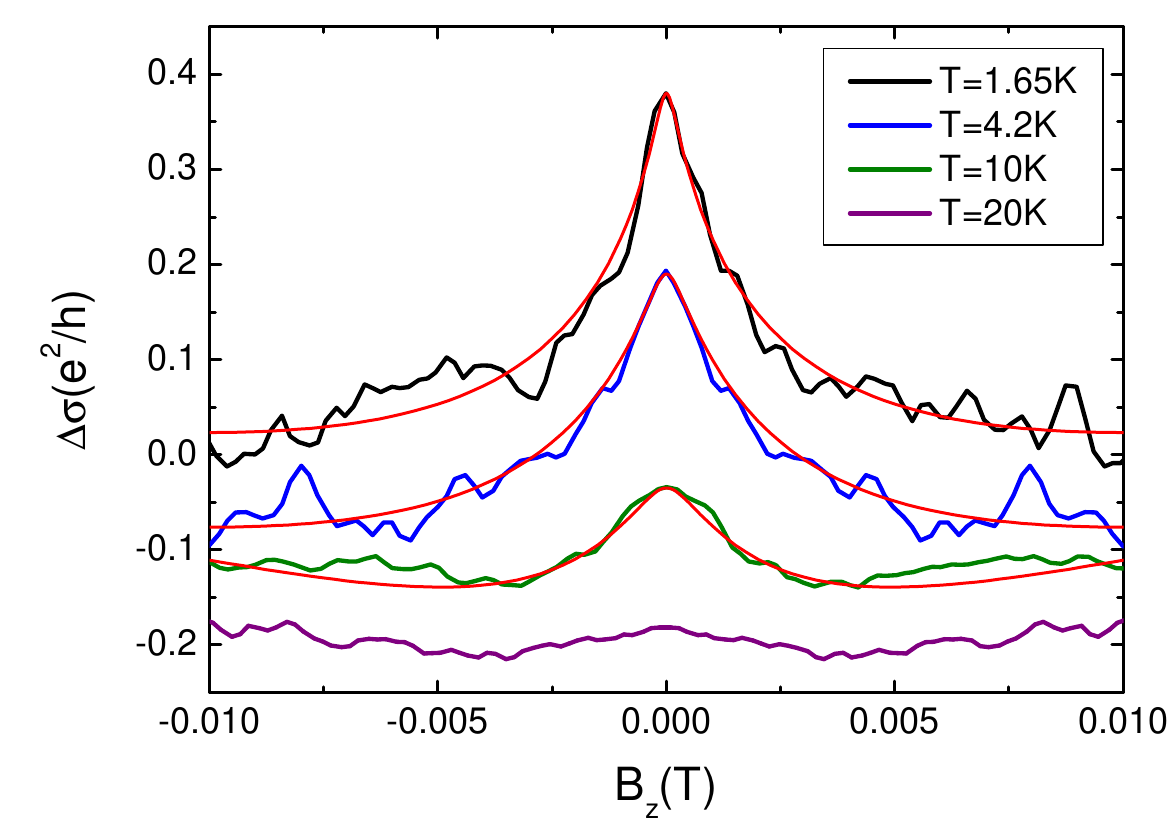}
	\caption{Gate averaged magnetoconductivity of a graphene/WSe$_2$ sample at different temperatures. The peak intensity decreases with increasing temperature as the phase coherence length decreases. The data was fitted by equation\,\ref{eq1}\,(red curves).\label{2}}
\end{figure}

For measurements in the diffusive regime monolayer graphene/WSe$_2$ is placed onto SiO$_2$ in device type\,1. We therefore observe a mobility of only $\mu=12\,000$\,cm$^2$/Vs at $T=1.65$\,K. Figure\,\ref{2} depicts the magnetoconductivity of this sample at different temperatures. In order to suppress universal conductance fluctuations an average over 15 curves at slightly different backgate voltages with a mean charge carrier concentration of $n=1.0 \cdot 10^{12}$/cm$^2$ was taken. The curves were obtained in a four-point lock-in measurement with an AC-current of $I_{AC}=10$\,nA for the curves at $T=1.65$\,K and $T=4.2$\,K, $I_{AC}=50$\,nA at T=10\,K and $I_{AC}=100$\,nA at T=100\,K at a frequency of f=13\,Hz.\\
The occurrence of a sharp peak in the magnetoconductivity can be explained by weak-antilocalization, stemming from spin-orbit coupling. For the case that the intervalley scattering rate exceeds the decoherence rate, the low magnetic field dependence of the conductivity correction, due to WAL can be described as\,\cite{McCann2012}:
\begin{equation}
\begin{split}
	\Delta \sigma (B)=-\frac{e^2}{2 \pi h} \Bigg[ F \left( \frac{\tau_B^{-1}}{\tau_\phi^{-1}} \right) -F \left( \frac{\tau_B^{-1}}{\tau_\phi^{-1}+2 \tau_{asy}^{-1}} \right)  \\
	-2F \left( \frac{\tau_B^{-1}}{\tau_\phi^{-1}+\tau_{so}^{-1}} \right) \Bigg]
\end{split}
\label{eq1}
\end{equation}
where $F(x)=\ln(x)+ \Psi(1/2+1/x)$, with $\Psi(x)$ being the digamma function, $\tau_B^{-1}=4DeB/\hbar$, $\tau_\phi$ the phase coherence time, $\tau_{so}$ the spin-orbit scattering time and $\tau_{asy}$ a scattering time that takes into account only spin-orbit coupling that is asymmetric in $z \rightarrow -z$ direction. Here $\tau_{so}$ combines symmetric and antisymmetric spin-orbit scattering: $\tau_{so}^{-1}=\tau_{sym}^{-1}+\tau_{asy}^{-1}$ \cite{McCann2012}.\\
Fitting the curve in Fig.\,\ref{2} at $T=1.65$\,K\,(red curves in Fig.\,\ref{2}) gives $\tau_\phi=25.7$\,ps, $\tau_{so}=0.57$\,ps and $\tau_{asy}=1.71$\,ps. These are comparable to the values that were reported for graphene placed on WSe$_2$\,\cite{Wang2016} and WS$_2$\,\cite{Wang2016,Wang2015a,Yang2016}. $\tau_{so}$, which is an upper bound for the spin-relaxation time is therefore much shorter than the values typically found in pristine graphene\,(100\,ps-1\,ns)\,\cite{Tombros2007,Han2010,Volmer2013}. The occurrence of WAL with such small $\tau_{so}$ is therefore a clear indication of strong SOC in this device. With increasing temperature the feature in Fig.\,\ref{2} decreases as the phase coherence time $\tau_\phi$ decreases and the peak disappears at $T=20$\,K.\\
The dual gated device allows us to examine the WAL peak with an applied transverse electric field, while leaving the charge carrier density unchanged.
\begin{figure}
		\includegraphics[width=\columnwidth]{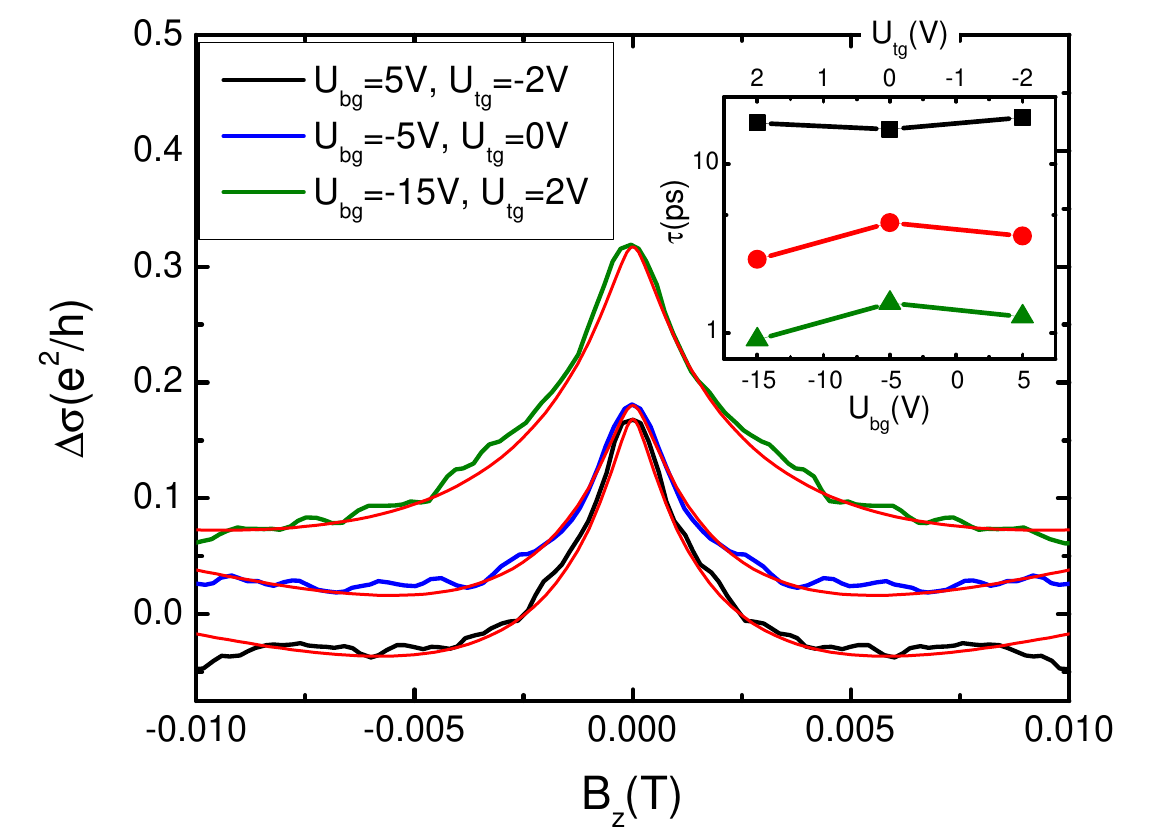}
	\caption{Gate averaged magnetoconductivity at three different top- and backgate voltage combinations. The top- and backgate voltages are chosen in a way that leaves the charge carrier concentration unchanged, while a transverse electric field is applied. The curves are fitted by equation\,\ref{eq1}\,(red curves). Inset: Electric field dependence of $\tau_\phi$\,(black squares), $\tau_{so}$\,(green triangles) and $\tau_{asy}$\,(red circles).\label{6}}
\end{figure}
Figure\,\ref{6} shows the magnetoconductivity at three different top- and backgate voltage combinations. Applying the electric field strongly decreases $\tau_{so}$ from $\tau_{so}=1.5$\,ps to $\tau_{so}=0.91$\,ps in one direction of the electric field and $\tau_{so}=1.25$\,ps in the other direction. The SOC strength is expected to increase with an electric field, due to the Rashba effect\,\cite{Bychkov1984}. However, $\tau_{so}$ depends on the total out-of-plane electric field acting on the carriers, which is composed of the externally applied field, as well as an internal  field, due to the WSe$_2$-graphene interface. The weak asymmetry in the external electric field therefore points to a small contribution of an internal field. This is in contrast to the findings of Yang {\em et al.} in graphene/WS$_2$ samples, who reported a linear dependence of the spin orbit scattering rate $\tau_{asy}$ with the applied electric field, while they assume the symmetric part of the scattering rate to be zero\,\cite{Yang2016}.\\
Spin relaxation is expected to be dominated by the Dyakonov-Perel mechanism. The SOC strength $\Delta_{DP}$ can be estimated by\,\cite{Huertas-Hernando2009}:
\begin{equation}
	 \tau_{so}^{-1}=4 \tau_e (\Delta_{DP}/\hbar)^2
\end{equation}
This results in a SOC strength of $\Delta_{DP}=0.7-1.0$\,meV, which agrees well with theoretical predictions\,\cite{Gmitra2016}. For the case of Elliot-Yafet dominated spin relaxation the SOC strength can be estimated by\,\cite{Huertas-Hernando2009}: $\tau_{so}^{-1}=\tau_e^{-1} \Delta_{EY}^2/E_F^2$. This results in an unrealistically large SOC strength of $\Delta_{EY}=35-65$\,meV. Further, we observe a decrease of $\tau_{so}$ with increasing charge carrier concentration, which indicates that spin relaxation is dominated by the Dyakonov-Perel mechanism.\\
\subsection{Ballistic Regime}
\begin{figure}
	\includegraphics[width=\columnwidth]{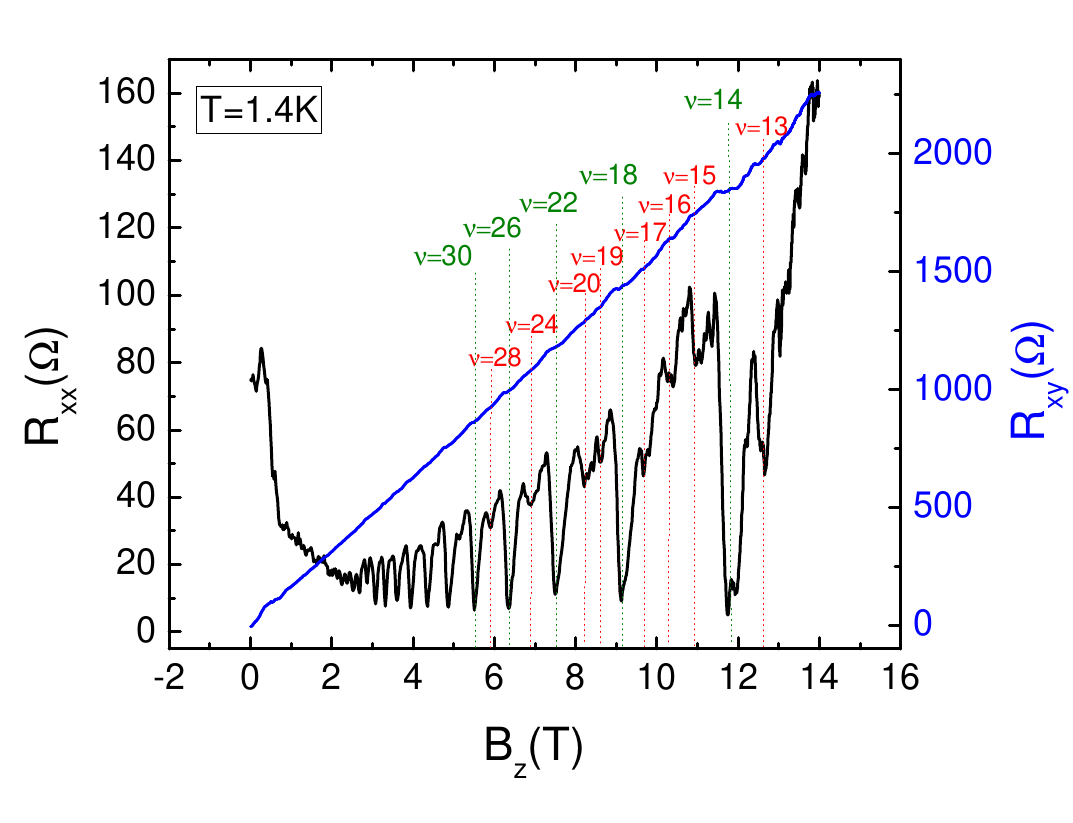}
	\caption{Shubnikov-de Haas oscillations\,(black curve) and quantum-Hall effect\,(blue curve) in hBN/graphene/WS$_2$. The splitting of valley and spin degeneracy in the Landau-levels indicates a high mobility of the sample.\label{3}}
\end{figure}

\begin{figure}
	\includegraphics[width=\columnwidth]{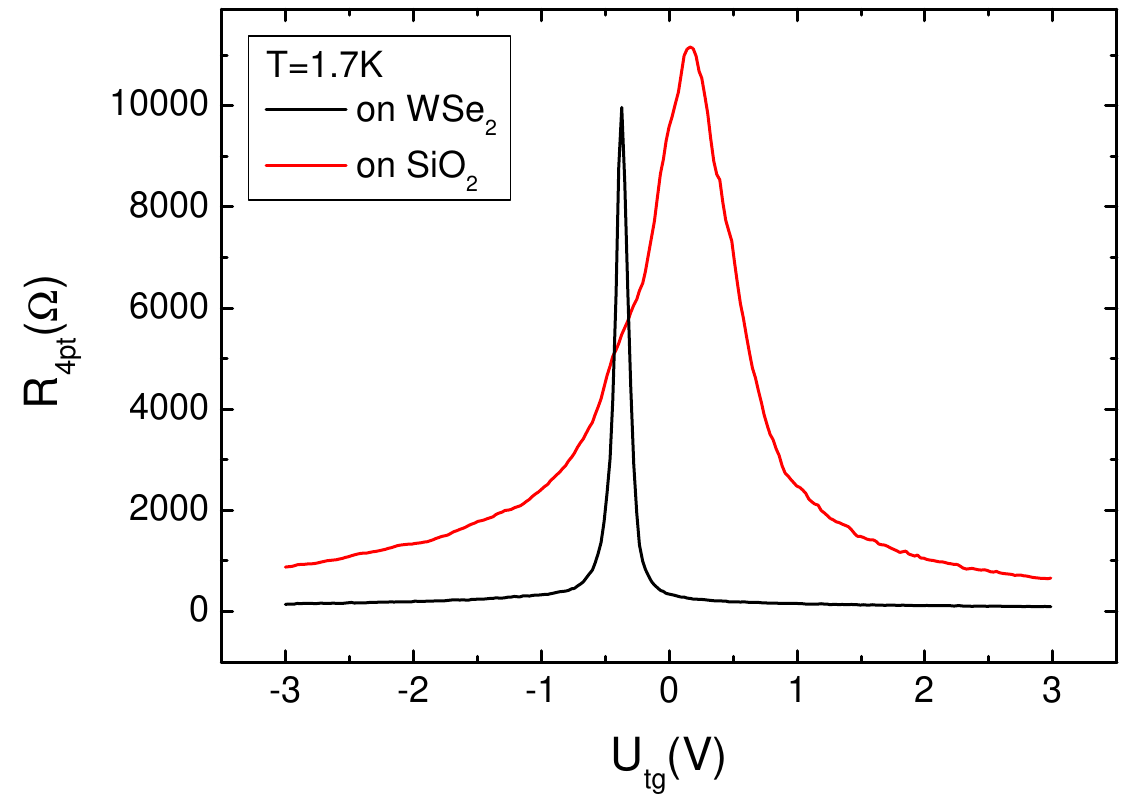}
	\caption{Topgate-sweeps of the sample depicted in \ref{1}\,(c). The black curves shows the resistance of the sample region lying on WSe$_2$, while the red curve depicts the resistance of the sample region on SiO$_2$.\label{4}}
\end{figure}

\begin{figure*}
	\includegraphics[width=1.6\columnwidth]{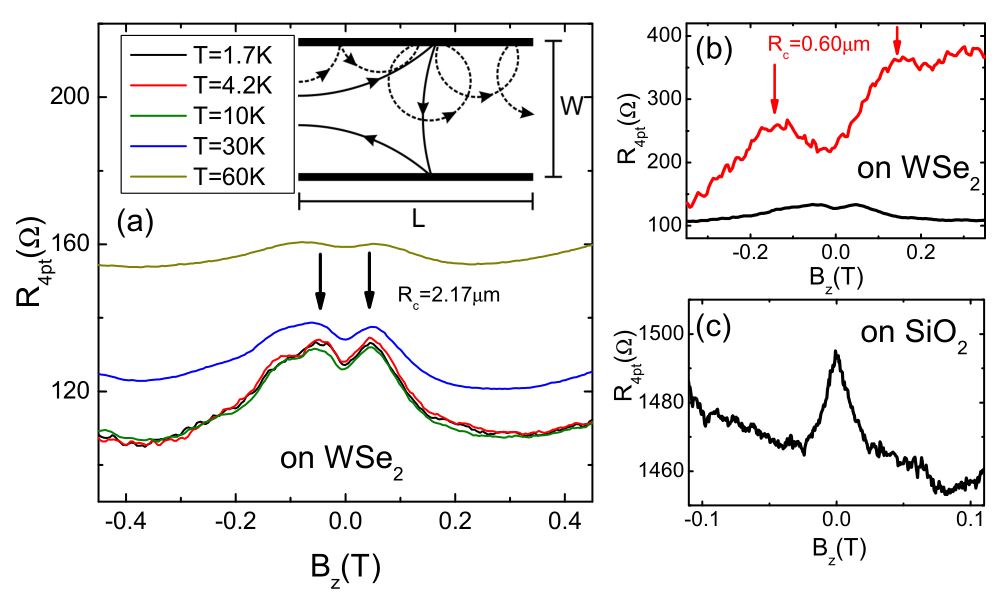}
	\caption{(a)\,Magnetoresistance of hBN/graphene/WSe$_2$ at different temperatures. Inset: Schematic electron trajectories for $R_C>W$\,(solid lines) and $R_C<W$\,(dashed lines). (b)\,Magnetoresistance for two samples with different width. The feature around $B=0$\,T is more pronounced and broader for a sample with width $W=1$\,$\mu$m\,(red line), than for a sample with width $W=4$\,$\mu$m\,(black line). This indicates that this feature is caused by a size effect, due to boundary scattering. Further, a linear background can be observed in the red curve. (c)\,Magnetoresistance of the sample area on SiO$_2$ shows weak localization behavior.\label{5}}
\end{figure*}

In order to increase the mobility of graphene we have encapsulated graphene between WSe$_2$\,(WS$_2$) and hBN\,(see Fig.\,\ref{1}\,(b)). Figure\,\ref{3} shows Shubnikov-de Haas oscillations\,(black curve) and quantum-Hall effect\,(blue curve) of device\,2, containing monolayer graphene between hBN and WS$_2$. 
This device showed mobilities of $\mu=50\,000$\,cm$^2$/Vs on the hole side and $\mu=120\,000$\,cm$^2$/Vs on the electron side. 
{In Fig. \ref{3}, a lifting of the spin and valley degeneracies can be observed, which results in integer filling factors in addition to the expected values of $4n+2$ for monolayer graphene. This behavior is typical for high mobility graphene\,\cite{Young2012}.}
The resistance peak at low magnetic field, followed by a negative magnetoresistance behavior will be discussed in the following sections.\\
In order to directly compare the substrates WSe$_2$ and SiO$_2$ in device 3, a bilayer graphene/hBN stack was placed in such a way that part of the stack lies on a WSe$_2$-flake and part of it lies directly on the SiO$_2$ substrate\,(see Fig.\,\ref{1}\,(c)). Figure\,\ref{4} shows topgate-sweeps of the four-point resistance of these two areas at $T=1.7$\,K.
From this we extract a mobility of $\mu=3200$\,cm$^2$/Vs on the hole side and $\mu=5300$\,cm$^2$/Vs on the electron side for the graphene on SiO$_2$. For the graphene on WSe$_2$ we extract $\mu=57\,000$\,cm$^2$/Vs and $\mu=92\,000$\,cm$^2$/Vs for hole and electron sides. The overall high mobilities resulting from encapsulation confirm the suitability of WS$_2$ and WSe$_2$ as substrates for high mobility graphene.\\
Figure\,\ref{5}\,(c) shows the magnetoresistance of the graphene on the SiO$_2$ substrate. 
Here we observe a peak in the resistance around $B=0$\,T, which we ascribe to weak localization. Fitting this peak with the formula for weak localization in bilayer graphene\,\cite{Gorbachev2007} reveals a phase coherence length of $L_\phi \sim 490$\,nm and an intervalley scattering length of $L_i \sim 420$\,nm.\\
For the part of the bilayer graphene on WSe$_2$ we observe a dip in the resistance around $B=0$\,T in Fig.\,\ref{5}\,(a). At first glance this feature might be interpreted as WAL. However, this dip is much too large\,($\Delta \sigma=20$\,e$^2$/h) and too broad to be fitted with equation\,\ref{eq1}. Further, the temperature dependence is much weaker and the dip is still visible at $T=60$\,K, in contrast to the WAL feature in Fig.\,\ref{2}.
Figure\,\ref{5}\,(b) shows the magnetoresistance of two bilayer graphene samples with different width. 
{While the black curve shows the magnetoresistance of the sample from Fig. \,\ref{4} and \ref{5}\,(a), with a width of $W=4\,\mu$m, the red curve shows the magnetoresistance of a sample with width $W=1.5\,\mu$m. The mobility of this sample was $\mu=90\,000$\,cm$^2/$Vs on the hole side and $\mu=100\,000$\,cm$^2/$Vs on the electron side.}
This behavior, i.e. the resistance peak at finite $B$, we ascribe to a ballistic effect, stemming from diffusive boundary scattering\,\cite{Beenakker1991,Thornton1989,Masubuchi2012}. A schematic description of this effect is shown in the inset of Fig.\,\ref{5}\,(a). At low magnetic fields the scattering between boundaries and therefore, the overall resistance, is initially increased\,(solid lines in the inset of Fig.\,\ref{5}\,(a)). 
{When the cyclotron diameter becomes smaller than the sample width, the scattering between boundaries is suppressed and therefore the resistance decreases\,(dashed lines in the inset of Fig.\,\ref{5}\,(a)).}
From the curves in Fig.\,\ref{5}\,\,(b), the cyclotron radius $R_c$ at the magnetic field, where the resistance reaches the maximum can be calculated as: 
\begin{equation}
R_c(B)=\frac{\hbar k_F}{e B_{max}}=\frac{\hbar\sqrt{\pi n}}{e B_{max}}.
\label{eq2}
\end{equation}
The calculated cyclotron radii are $R_c=2.17\,\mu$m for the sample with width $W=4\,\mu$m and $R_c=0.60\,\mu$m for the sample with width $W=1\,\mu$m. 
This shows that $R_c$ scales with the sample width $W$. For semiconductor 2DEGs, a relation $W=0.55 R_c$ was found \cite{Thornton1989}, whereas for 
hBN encapsulated graphene a different prefactor was observed \cite{Masubuchi2012}.
{The resistance peak at low magnetic field in Fig.\,\ref{3} is also attributed to this effect.}\\
No WAL behavior could be observed for graphene encapsulated between hBN and WSe$_2$\,(WS$_2$). 
We attribute this to a transition from the diffusive to the quasiballistic regime.
{Since, equation\,\ref{eq1} was developed in the diffusive regime, it is only valid for the case of: $\tau_{\phi}>\tau_{asy}>\tau_{so}>\tau_e$. Due to the higher mobility for decives of type\,2 and 3, we find $\tau_e$ to be in the range of $\tau_e \approx 1\,$ps. Therefore the relation $\tau_{so}>\tau_e$ may not be valid here. We expect WAL to be suppressed, due to reduced backscattering and the WAL peak to be narrower, resulting from the higher mobility in these samples\,(a similar behavior has been observed in GaAs heterostructures\,\cite{Grbiifmmodeacutecelsecfi2008}). Therefore the absence of WAL in these samples is not indicative of a lower SOC strength.}\\
\section{Conclusion}
In conclusion we investigated charge transport in several graphene/WSe$_2$\,(WS$_2$) heterostructures. We successfully employed the established fabrication techniques for hBN/graphene/hBN stacks to heterostructures containing WSe$_2$\,(WS$_2$) and graphene. Placing a graphene/WSe$_2$ stack on SiO$_2$ resulted in a mobility of $\mu=12\,000$\,cm$^2$/Vs. In this sample we observed a peak in the magnetoconductivity, which we attributed to the WAL effect, stemming from SOC. Applying an electric field increased the SOC strength in this sample. Encapsulating graphene between WSe$_2$\,(WS$_2$) and hBN increased the mobility to up to $\mu=120\,000$\,cm$^2$/Vs. No WAL behavior could be observed in these samples. We attribute this to a transition from the diffusive to the quasiballistic regime. This is further confirmed by the occurrence of a quasiballistic size effect, due to diffusive boundary scattering. These results confirm the suitability of WSe$_2$\,(WS$_2$) as a substrate for high quality graphene with strongly increased SOC.

\begin{acknowledgments}
Financial support by the Deutsche Forschungsgemeinschaft (DFG) within the programs GRK 1570 and SFB 689 is gratefully acknowledged. The authors would like to thank J. Fabian and T. Korn for fruitful discussions.
\end{acknowledgments}

\end{document}